\documentclass[11pt,a4paper]{article}
\usepackage{authblk}
\usepackage{cite}
\usepackage{amsmath}
\usepackage{amssymb}
\usepackage{geometry}
\title{The zero mass limit of Kerr and Kerr-(anti-)de-Sitter space-times: exact solutions and wormholes}

\author[1]{T. Birkandan \footnote{E-mail: birkandant@itu.edu.tr}}
\affil[1]{Istanbul Technical University, Department of Physics, Istanbul, Turkey.}

\author[2]{M. Horta\c{c}su \footnote{E-mail: hortacsu@itu.edu.tr}}
\affil[2]{Mimar Sinan Fine Arts University, Department of Physics, Istanbul, Turkey.}

\begin{document}
\maketitle

\begin{abstract}
Heun-type exact solutions emerge for both the radial and the angular equations for the case of a scalar particle coupled 
to the zero mass limit of both the Kerr and Kerr-(anti)de-Sitter spacetime. Since any type D metric has Heun-type solutions, 
it is interesting that this property is retained in the zero mass case. This work further refutes the claims that $M$ going to zero limit of the Kerr metric  is
 both locally and globally the same as the Minkowski metric.
\end{abstract}





\section{Introduction}

In a very interesting recent paper \cite{Gibbons}, Gibbons and
Volkov studied the zero mass limit of the Kerr \cite{Kerr} metric
to show that the zero mass limit of this metric is a wormhole, and
is not just the Minkowski space metric as stated by some authors
 (see for example \cite{Landau,Visser}). This Gibbons-Volkov paper is a
continuation of research  where they  showed that similar behavior
exists in many other metrics, namely Schwarzschild \cite{Volkov1}
and Weyl metrics \cite{Volkov2}. They also  show that the Klein-Gordon equation, in  in spheroidal
coordinates, written in the background of the Kerr-(anti) de Sitter metric, allows  seperation of variables, resulting in two equations for the radial and angular($\theta$) variables. Here we obtain the exact   solutions of the wave equation of a scalar particle with mass $\mu$ in the background of these metrics. 
 
 \section{ Limit of the Kerr metric}
  In the first paper cited above
\cite{Gibbons}, they first write the Minkowski metric in
cylindrical coordinates
\begin{equation}
 ds^2 = -dt^2 + d\rho^2 + \rho^2 d\phi^2 +dz^2,
\end{equation}
and then transform to oblate spheroidal coordinates,
\begin{equation}
ds^2 = -dt^2 + {\frac{r^2+ a^2 \cos^2{\theta}}{ r^2 + a^2}} 
\bigg(dr^2+( r^2 + a^2) \cos^2 {\theta} \bigg) + (r^2 +a^2) sin^2{\theta } d\phi^2.
\end{equation}
In the transformation to oblate spheroidal coordinates,
$\rho=(r^2 + a^2)^{1/2}\sin{\theta}$, 
$z =r \cos{\theta}$,  $\rho $ is not the same coordinate as used in the
cylindrical coordinates.

The authors of \cite{Gibbons} note that ``the Jacobean of the second
transformation is zero for $\rho = x^2 + y^2 = a^2, z=0 $, which
is a ring of radius $a$ in the equatorial plane" \cite { Gibbons}.
Gibbons et al \cite{Gibbons} state that one obtains  the metric given above 
in terms of the
 oblate spheroidal coordinates also in the
zero mass limit of the Kerr spacetime \cite{Carter}.   
 One can
separate the variables in spheroidal coordinates by assuming a
solution of the form $\Phi = e^{i\omega t+im \phi}F(r) G(\theta)$.
Then the Klein-Gordon equation reduces to  
radial and angular equations:
\begin{eqnarray}
&&\frac{d}{dr} \bigg((r^{2} + a^{2} ) \frac{dF(r)}{dr}\bigg)  
+\bigg( r^{2} \omega^2 -\mu^{2} r^{2}+\frac{m^{2} a^{2} }{ r^{2} + a^{2}} + \lambda \bigg) F(r) =0,
\end{eqnarray}
\begin{eqnarray}
&&\frac{1}{\sin\theta} \frac{d}{d \theta}\bigg(\sin{\theta}  \frac{dG (\theta)}{d \theta} \bigg) 
+\bigg( a^{2} \cos^{2} \omega^2 - \mu^{2} a^{2} \cos^{2} {\theta}  - \frac{ m^{2}  }{\sin^{2}\theta} - \lambda \bigg)G(\theta) = 0.
\end{eqnarray}
 The independent variable $r$ appears in the  radial equation with even powers. 
Thus any solution which is valid for positive $r$ is also valid 
for its negative values.

For the angular variable $\theta$, we make the transformation $x= \cos^{2}\theta$ 
which changes the angular equation to
\begin{eqnarray}
&&\frac{{d^{2}G (x)}}{{d x^{2}}}+\bigg( \frac{1}{x-1} 
 +\frac {1}{2x} \bigg) {\frac{{dG (x) }}{{d x }}} \nonumber\\
&&+{\frac{{1}}{{4x(x-1)
 }}} \bigg(  \mu^{2} a^{2} x -a^{2}\omega^{2} 
- \frac{m^2 }{x-1} - \lambda \bigg) G(x) = 0.
\end{eqnarray}
This equation has a  solution in terms of a confluent Heun function ($H_C$) 
multiplied by some expressions with powers \cite{Heun,Ronveaux,Slavyanov,Hortacsu}
\begin{equation}
G(x) = x^{1/2} (x-1)^{m/2}  H_C (\alpha,\beta,\gamma,\delta, \eta;z),
\end{equation}
where the parameters of the confluent Heun function are given by
\begin{eqnarray}
\alpha  &=& 0, \\
\beta &=& -\frac{1}{2}, \\
\gamma&=& \frac{m}{2}, \\
\delta &=& -\frac{{1}}{{4(\omega^{2} + \mu^{2}) a^{2}}} , \\
\eta &=& -\frac{{1}}{{4(m^{2} + \lambda +1)}}, \\
z&=&\cos^{2}\theta.
\end{eqnarray}
Here we used the form of the confluent Heun equation given by Fiziev \cite{Fiziev1, Fiziev2}. After expressing $G$  by a function $H$ multiplied by the terms given above, we matched  the resulting equation for $H_C$ to the form below.
\begin{equation}
    {\frac{{d^{2}H_C}}{{dz^{2}}}}+\left( \alpha +{\frac{{\gamma+1}}{{z-1}}}+{\frac{{\beta+1}}{{z}}}%
    \right) {\frac{{dH_C}}{{dz}}}+\left({\frac{{\mu}}{{z}}}+ {\frac{{\nu}}{{z-1}}}\right)H_C =0, \label{fizheun}
\end{equation}
with solution $H_C(\alpha, \beta, \gamma, \delta,\eta,z)$, and the parameters $\delta$ and $\eta$ are given by  the relations
\begin{equation}
    \delta = \mu+\nu-\alpha \bigg( \frac{{ \beta+\gamma+2}}{{2}} \bigg),
\end{equation}
\begin{equation}
    \eta = \frac{{ \alpha(\beta+1) }}{{2}} - \mu - \frac {{ \beta+\eta+ \beta \gamma}}{{2}}.
\end{equation}
The  regular singular points are at $x=0, 1$ and the irregular singular point is at infinity. 

We think to obtain a confluent Heun solution even at the limit when $M$ goes to zero is interesting, since it is known that the wave equation in the background of the Kerr metric results in a  confluent Heun solution \cite{ Blaudin,Leaver,Suzuki,Batic}. Here we see that this characteristic is retained in the mass going 
to zero limit.

We also tried to see if there are polynomial solutions which are analytic at both of the regular singularities in this case of the  confluent Heun equation. One of the criteria to have such a solution is given in \cite{ciftci, karayer}  which requires  the parameter $\alpha$ not equal to zero, which is not true in our case. Our solution is analytic only in a space where the $z$-axis is excised. This validates the result of the Gibbons and Volkov paper \cite{Gibbons}.

We were surprised  to find that if we make the variable transformation $ u=-r^2/a^2 $, we get exactly the same expression 
for the radial equation,
 given above, as the angular
equation. The solution is exactly the same with the appropriate variable change. Now the singularity at $u=1$ is not represented on 
the real space.
 Here we explicitly see the square root cut at $r=0$, which is $\rho= a $, as noted by Gibbons and Volkov \cite{Gibbons}, which increases the range of the angle
 $\phi$ two fold, verifying their result. 

\section{Limit in the Kerr-de Sitter metric}
At the end of the first paper cited above, Gibbons and Volkov also give an example of a scalar particle coupled
to the wave equation obtained in the zero mass limit of the Kerr-(anti)de Sitter
(Kerr-(A)dS) spacetime \cite{Carter}. They can separate the
variables in spheroidal coordinates by assuming a solution of the
form $\Phi = e^{i\omega t+im \phi}F(r) G(\theta)$. Then the radial and angular parts of the Klein-Gordon equation reduce to
\begin{eqnarray}
&&\frac{d}{dr} \bigg((r^{2} + a^{2} ) D \frac{dF(r)}{dr}\bigg)  
+ \bigg( \frac{3 \Xi \omega^2}{\Lambda D} -\mu^{2} r^{2}+\frac{m^{2} a^{2} \Xi}{ r^{2} + a^{2}} + \lambda \bigg) F(r) =0,
\end{eqnarray}
\begin{eqnarray}
&&\frac{1}{\sin\theta} \frac{d}{d \theta}\bigg(\sin{\theta} C \frac{dG (\theta)}{d \theta} \bigg) 
+\bigg(\frac{3 \Xi\omega^2}{\Lambda C} + \mu^{2} a^{2} \cos^{2} {\theta}  + \frac{ m^{2} \Xi }{\sin^{2}\theta}
 + \lambda \bigg)G(\theta) = 0.
\end{eqnarray}
Here $ D= 1- {\frac{{ \Lambda r^{2} }}{{3}}}$, $C= 1+ {\frac{{ \Lambda a^{2} }}{{3}}\cos^2\theta}$ and
$\Xi= 1 + {\frac{{\Lambda a^2 }}{{3}}}$. The independent variable $r$ appears in the  radial equation with even powers. Thus any solution which is valid for positive $r$ is also valid for its negative values.

For the angular variable $\theta$, we again make the transformation $x= \cos^{2}\theta$ which changes this equation to
\begin{eqnarray}
&&\frac{{d^{2}G (x)}}{{d x^{2}}} +\bigg( \frac{1}{x-1} + \frac{1}{x+{\frac{3}{a^{2}
 \Lambda}}}+{\frac {{1}}{{2x}}} \bigg) {\frac{{dG (x) }}{{d x }}} \nonumber\\
&&+{\frac{{1}}{{x(x-1)(x+{\frac{{3}}
 {{a^2 \Lambda}}})}}} \bigg( \frac{{3 (1+{\frac{{3}}{{a^2 \Lambda }}}) \omega^{2}}}{{\Lambda (x+{\frac{{3}}{{a^2 \Lambda }}}})}
  + \mu^{2} a^{2} x 
+ \frac{m^2 (1+\frac{3}{a^2 \Lambda})}{x-1} + \lambda \bigg) G(x) = 0.
\end{eqnarray}
This equation has a  solution in terms of a general Heun function ($H_G$) multiplied by some expressions with powers \cite{Heun,Ronveaux,Slavyanov,Hortacsu}
\begin{equation}
G(x) = x^{1/2} (x-1)^{m/2} \big(x+{\frac{{3}}{{a^2 \Lambda}}}\big)^{\frac{i\omega}{2} \sqrt{\Lambda/3}}
 H_G (a,q,\alpha,\beta,\gamma,\delta;x),
\end{equation}
where the parameters of the general Heun function are given by
\begin{eqnarray}
a&=& -{\frac{{3}}{{a^2 \Lambda}}}, \\
-q&=&-\frac{i\omega \sqrt{3/\Lambda}}{4} + \frac{3m}{4a^2 \Lambda}+\frac{m^2}{4} (\frac{3}{a^2 \Lambda}+1)  
+ \frac{3\lambda}{4a^2\Lambda}+\frac{3\omega^2}{4\Lambda}(\frac{3}{a^2 \Lambda}+1), \\
\alpha  &=& \frac{1}{2}( i\omega \sqrt{3/\Lambda} +m+\frac{3}{2}) + \sqrt{ \frac{9}{4} - \frac{3\mu^2}{\Lambda}}, \\
\beta &=& \frac{1}{2}( i\omega \sqrt{3/\Lambda} +m+\frac{3}{2}) - \sqrt{ \frac{9}{4} - \frac{3\mu^2}{\Lambda}}, \\
\gamma&=& \frac{1}{2}, \\
\delta &=& m+1, \\
x&=&\cos^{2}\theta.
\end{eqnarray}
Here we matched the parameters to the standard Heun equation \cite{Arscott}
\begin{eqnarray}
&&\frac{{d^{2}H_G (x)}}{{d x^{2}}} +\bigg( \frac{\delta}{x-1} + \frac{\epsilon}{x-a}+ \frac{{\gamma}}{{x}} \bigg) {\frac{{dH_G (x) }}{{d x }}} 
+\bigg(\frac{{\alpha \beta x-q}}{{x(x-1)(x-a)}} \bigg) H_G(x) = 0.
\end{eqnarray}
This equation has also the constraint $ \alpha+\beta+1= \gamma +\delta+\epsilon $ on the parameters if we want the behavior at infinity 
as $ z^{-\alpha} $ or $ z^{-\beta}$ plus terms with 
higher powers of $ 1/z $.
In our case the singular point at $x= -{\frac{{3}}{{a^2 \Lambda}}}$ is only for imaginary values of the variable $\cos\theta$, 
which is not in physical domain of interest. 

We think finding again a Heun type solution is interesting, since it is known that any wave equation in the background 
of a type D metric results in a Heun class solution \cite{ Blaudin,Leaver,Suzuki,Batic}. 
Here we see that  again this characteristic is retained in the mass going to zero limit.

We also tried to see if there are polynomial solutions to have a Heun solution which is analytic at both 
of the remaining singularities, at $x=0$ and $x=1$. The criteria to have such a solution are given by Arscott \cite{arscott1}. we see that we cannot satisfy these conditions even if we set the mass of the scalar particle $\mu$ equal to zero. Our solution is analytic only in a space where the $z$-axis is excised. This validates the result 
of the Gibbons and Volkov paper \cite{Gibbons}.

We also  found that if we make the variable transformation $ u= - r^2/a^2 $, we again get exactly the same expression 
for the radial equation, given above, as the angular
equation. The solution is exactly the same with the appropriate variable change. Now the singularity at $u=1$ is not represented 
on the real space. Here we explicitly see the square root cut at $r=0$ noted by Gibbons and Volkov \cite{Gibbons}, 
which increases the range of the angle $\phi$ two fold, verifying their result. For the radial equation  the singularity at
 $u=-r^{2}/a^{2}= -{\frac{{3}}{{a^2 \Lambda}}}$ is in the physical domain, and corresponds to the event horizon singularity. 

We changed our variable to $v= u+{\frac{{3}}{{a^2 \Lambda}}}$ and investigated the behavior beyond the event horizon. The equation reads
\small\begin{eqnarray}
&& \frac{d^{2}G(v)}{d v^{2}}+\bigg(\frac{1}{v-\frac{3}{a^{2}\Lambda}-1} + {\frac{{1}}{{v}}}
+ {\frac {{1}}{{2({{v-{\frac{{3}}{{a^{2}\Lambda}}}}})}}} \bigg){\frac{{dG (v) }}{{d v }} } \nonumber \\
&& + {\frac{{1}}{{v(v-1-{\frac{{3}}{{a^{2}
\Lambda}}})(v-{\frac{{3}}{{a^2 \Lambda}}})}}} \bigg(
{\frac{{(1+{\frac{{3}}{{a^2 \Lambda }}}) \frac{3}{4}
\omega^2}}{{\Lambda (v-{\frac{{3}}{{a^2 \Lambda }}}) }}} \nonumber \\
&&+ \mu^{2}a^{2} (v+{\frac{{3}}{{a^2 \Lambda}} )}- {\frac{{ m^{2}
(1+{\frac{{3}}{{a^2\Lambda}}})}}{{(v-1-{\frac{{3}}{{a^{2}
\Lambda}}})}}} + \lambda \bigg)G(v) = 0.
\end{eqnarray}
\normalsize
The solution of this equation is very similar to the one given above. We get another Heun type solution.

\section{Conclusion}
Here, we verify the results of the Gibbons and Volkov paper \cite{Gibbons} by solving the wave equation for a scalar particle in the $M$ going to zero limit of both the Kerr and Kerr-(anti)de Sitter metrics. Using the 
properties of the  solution of the scalar 
wave equation, we show that the zero mass limit of Kerr spacetime indeed is not just a Minkowski space. 
The non analyticity (square root cut) and excising of the $z$-axis are evidences of the fact that globally this space is 
different from the Minkowski space.

\section*{Acknowledgement}  
M.H. thanks O\u{g}uzhan Ka\c{s}{\i}k\c{c}{\i} for technical assistance. He also thanks the Science Academy, Istanbul for support. This work is supported by T\"{U}B\.{I}TAK, the Scientific and Technological Council of Turkey.

\end{document}